\documentclass[sigplan,nonacm]{acmart}

\usepackage{siunitx}
\usepackage{amsmath}
\usepackage{courier}
\usepackage{comment}
\usepackage[scaled]{helvet}
\usepackage{url}
\usepackage{listings}
\usepackage{enumitem}
\usepackage{xspace}
\usepackage{graphicx}

\definecolor{CBlue}{HTML}{5B9BD5}
\definecolor{COrange}{HTML}{ED7D31}
\definecolor{CGreen}{HTML}{70AD47}
\definecolor{CRed}{HTML}{FF6B6B}
\definecolor{CGray}{HTML}{A9A9A9}
\definecolor{CPurple}{HTML}{7B68EE}
\definecolor{CTeal}{HTML}{2EC4B6}
\definecolor{CDkBlue}{HTML}{2F5597}

\newcommand{\sys}{\textsc{fabric\_ext}\xspace}

\newcommand{\xiangyu}[1]{}

  {\begin{enumerate}[itemsep=-0pt, parsep=0pt, topsep=0pt, leftmargin=1pc]}
  {\end{enumerate}}

\begin{document}

\title{The Fabric Is the Cluster Driver: Cross-Layer eBPF Policies for GPU--CXL Fabrics}

\author{Yiwei Yang}
\affiliation{
  \institution{UC Santa Cruz}
  \country{USA}
}
\email{yyang363@ucsc.edu}






\author{Andi Quinn}
\affiliation{
  \institution{UC Santa Cruz}
  \country{USA}
}
\email{aquinn1@ucsc.edu}


\begin{abstract}
Modern GPU clusters are no longer only collections of accelerators behind a host CPU.
They are fabrics: GPU kernels expose execution phase, warp behavior, page faults, KV blocks, and tensor semantics; NICs and DPUs expose queue pairs, work queues, completions, retransmits, and congestion; CXL switches and memory devices expose placement, sharing, ordering, persistence, and near-memory movement.
Yet today's control planes split these signals across mutually blind policy systems.
The GPU scheduler does not know which RDMA queue carries decode KV fetches, and the network or CXL fabric does not know whether a transfer belongs to prefill, decode, MoE routing, checkpointing, or background training.

We present \sys, an eBPF middleware compiler and runtime for extensible OS policies over GPU--CXL fabrics.
\sys lets one policy program execute across GPU hooks, driver/runtime hooks, DPU/NIC hooks, and CXL switch or near-memory hooks.
The key abstraction is a semantic movement graph: edges describe bytes, stride, reuse distance, read/write ratio, source and destination, ordering requirement, alias set, ownership, and transformations such as Move, Quantize, Compress, Checksum, Filter, Reduce, Scatter/Gather, Replicate, and Persist.
The compiler lowers this graph into per-device eBPF programs, verifier obligations, consistency-classed BPF maps, and artifacts for dputime.
At the fabric edge, \sys treats a near-Type-2 small core as a hardware-JIT and state manager: it specializes verified movement descriptors into local copy, placement, ordering, and transformation commands, while the surrounding Von Neumann island of memory, DMA, and compute engines performs the dataflow.
Because this dataflow is data-driven, \sys also places observation beside the island, where queues, DMA completions, memory placement, and ownership transitions are visible as they happen.
The canonical stress case is LLM prefill: attention streams KV blocks and reductions while FFN streams activations, weights, and compressible intermediates, forcing one request to cross GPU tensor execution, DPU/NIC event execution, and CXL or switch-local dataflow islands.

\sys's central contribution is not a shared map.
It is a cross-device policy contract.
Telemetry maps may be eventually consistent; scheduling hints use bounded-staleness epochs; ownership and placement maps require atomic epoch handoff; command maps require exactly-once completion.
A cross-device effect verifier checks both SIMT GPU execution and event-driven fabric execution: bounded helpers and RDMA/CXL operations, no waits on completions triggered by the same handler, no GPU--fabric dependency cycles, tenant/QP/region capabilities, retry TTLs, and atomic policy updates across devices.

We prototype \sys in Damer with dputime as the eBPF frontend.
On 1{,}000 generated movement microbenchmarks, the compiler produces verified plans and BPF C in microseconds.
Across 14 LLM fabric workloads---including Qwen 27B decode, long-context CXL KV cache, pipeline and tensor parallelism, LoRA adapter updates, Qwen scale-out from 2 to 32 GPUs, MoE all-to-all, prefill/decode colocation, remote KV cache, and training/checkpoint traffic---\sys compiles 86 events into BPF objects and validates remote Arm compilation on a BlueField DPU.
The expanded suite reports a modeled 2.34$\times$ data-movement E2E speedup over staged host/NIC control.
A copy-path optimizer that selects DPU inline, GPU pre-transform, and direct GPU-peer/CXL paths improves the modeled speedup to 3.47$\times$.
These results are not token-latency claims; they show that semantic, verified fabric policies can expose where end-to-end data movement improves when the fabric, not the host CPU alone, becomes the cluster driver.
\end{abstract}

\maketitle

\section{Introduction}
\label{sec:intro}
GPU clusters are becoming programmable fabrics rather than host-attached islands.
An LLM serving request may start as prefill on one GPU, decode on another, fetch remote KV blocks from CXL memory, route tokens through MoE experts over RDMA, and periodically share the same fabric with background training and checkpoint traffic.
Each layer sees a useful part of the truth.
GPU kernels and runtimes see execution phase, streams, warps, page faults, KV block identifiers, expert IDs, and whether a token is on the latency-critical path.
NICs, DPUs, and switches see queue pairs, WQEs, CQEs, retransmissions, congestion, flow timeouts, and placement pressure.
CXL switches and memory devices see fabric routes, shared memory windows, persistence domains, and ordering constraints.
No single layer can currently express a policy that combines these facts.

This split causes a control-plane mismatch.
A GPU policy can decide that decode KV fetches are urgent, but it cannot directly pace a congested RDMA queue or redirect a transfer away from an overloaded expert replica.
A DPU or NIC policy can prioritize a QP or react to congestion, but it cannot tell whether the bytes are decode, prefill, MoE tokens, gradients, checkpoint records, or CXL-memory migration.
A CXL fabric manager can configure memory windows and routing, but it lacks application semantics such as reuse distance, alias ownership, or the phase in which a page fault occurs.
The result is a set of local controllers that each make reasonable decisions with incomplete information.

\sys takes the opposite view: the policy boundary should be the fabric, not the device.
It extends the eBPF programming model from one GPU or one network endpoint to a cross-device policy system spanning GPU kernels, GPU drivers, DPU/NIC execution, CXL switches, and CXL-attached memory.
The goal is not to replace RDMA, GPUDirect, CXL fabric management, or existing GPU runtimes.
Those mechanisms already provide high-performance data paths.
\sys contributes the missing safe, dynamic, semantic policy layer that decides \emph{what} movement should happen, \emph{where} transformations should execute, and \emph{when} fabric operations should be delayed, redirected, replicated, or persisted.
For CXL and switch-attached execution, the target is a near-Type-2 small core rather than a full host CPU.
That core acts as a hardware-JIT and state manager: it consumes verified movement descriptors, maintains epochs and ownership, emits local commands, and lets the adjacent Von Neumann island of memory, DMA, and simple compute engines carry out the dataflow.
Since that dataflow is driven by queue entries, memory completions, and buffer readiness rather than host polling, \sys places observation beside the island as well.
The same fabric-side runtime that programs movement can observe queue pressure, completion timing, ownership transitions, and local placement state at the point where they affect data movement.

The strongest use case is LLM prefill rather than ordinary QoS.
In attention, one request streams KV blocks, page-table state, reductions, and remote-cache hits.
In FFN, the same request streams activations, weights, quantized or compressed intermediates, and tensor-parallel exchanges.
These movements are data-driven and phase-dependent, so the right policy point changes within a single request.
GPU hooks see layer, phase, KV block, and tensor-shard semantics.
DPU/NIC hooks see RDMA pressure, queue completion, congestion, and remote GPU readiness.
A near-Type-2 fabric core sees CXL placement, local queues, DMA completion, and ownership transitions.
\sys lets one verified policy span all three architectures instead of approximating the request as a QP priority.

\begin{figure}[t]
    \centering
    \includegraphics[width=\columnwidth]{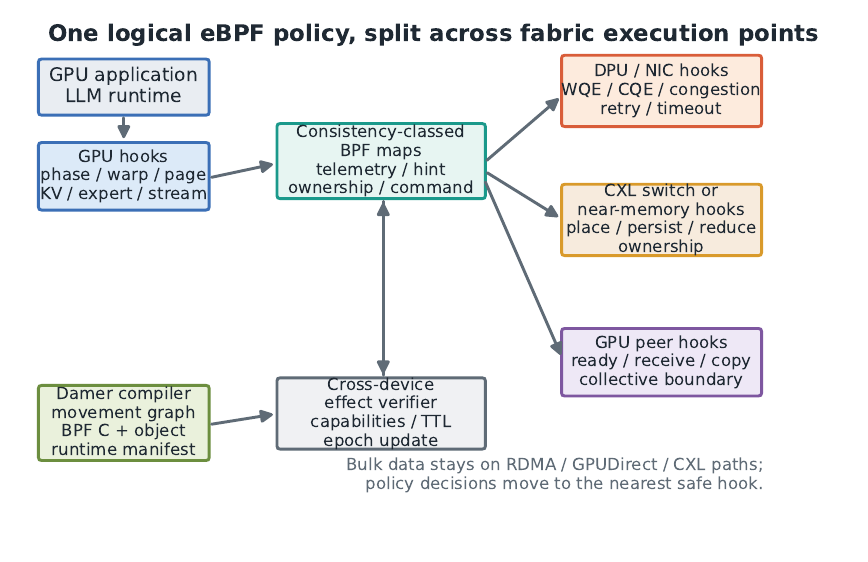}
    \caption{\sys compiles one logical eBPF policy into GPU, driver, DPU/NIC, and CXL switch or near-memory handlers. Bulk data continues to use RDMA, GPUDirect, and CXL data paths; \sys provides the verified semantic policy layer above them.}
    \label{fig:fabric-arch}
\end{figure}

Consider semantic RDMA scheduling for MoE inference.
Today the fabric sees packets or QPs.
The GPU sees expert IDs, token batches, and which experts are on the decode critical path.
With \sys, a GPU hook publishes a fabric hint keyed by stream, rank, and expert.
A fabric hook on a DPU, NIC, or CXL switch-attached compute engine uses that hint when WQEs are submitted or completions arrive: critical expert tokens are scheduled first, traffic to congested experts is paced, idle replicas receive redirected tokens, and transfers to a destination GPU that is not ready can be deferred before they build up in CQ or GPU buffers.
The same policy can run on the best available execution point: a BlueField DPA handler for RDMA queue decisions, a host dputime handler for driver-level fallback, or a switch/near-memory handler for CXL placement and persistence decisions.

The abstraction \sys exposes is a semantic data-movement graph.
Nodes include host memory, CXL memory devices, GPU memory, accelerators, and switch compute engines.
Edges carry bytes, stride, reuse distance, read/write ratio, source and destination, ordering requirement, alias set, ownership, and a transformation.
\sys supports fused actions that recur in GPU-fabric workloads:
Move, Quantize, Compress, Checksum, Filter, Reduce, Scatter/Gather, Replicate, and Persist, always represented as movement plus an optional transformation.
These are not packet classifiers.
They are movement intents that the compiler can place on GPUs, DPUs, CXL memory devices, or switch-local engines.

The hard part is safety.
Extending eBPF across the fabric cannot simply share maps between devices.
Prior GPU eBPF systems showed that CPU--GPU hierarchical maps are useful for statistics and hints, but they mostly tolerate eventual consistency.
Fabric policies also manipulate queueing, placement, ownership, and retries.
A stale statistic is acceptable; a stale ownership transfer or unbounded retry loop is not.
\sys therefore introduces consistency-classed BPF maps and a cross-device effect verifier.
Telemetry maps are eventually consistent.
Hint maps carry bounded-staleness epochs.
Ownership and placement maps use epoch-atomic handoff.
Command maps expose exactly-once completion.
The verifier checks both GPU SIMT constraints and fabric-handler constraints, including bounded helper calls, bounded RDMA/CXL operations per event, no dependency cycles across GPU and fabric programs, tenant/QP/region capabilities, and retry TTLs.

This paper makes five contributions:
\begin{itemize}[nosep]
\item We identify the \emph{semantic split} in GPU fabrics: GPUs know execution meaning but not fabric state; DPUs, NICs, and CXL switches know fabric state but not execution meaning.

\item We design \sys, a cross-device eBPF policy system that lowers semantic movement graphs into GPU, driver, DPU/NIC, and CXL-switch policy hooks.

\item We introduce a near-Type-2 fabric execution model in which a small core performs hardware-JIT specialization, policy state management, and adjacent observation while the surrounding memory/compute island executes data-driven dataflow.

\item We introduce consistency-classed BPF maps and a cross-device effect verifier for policies whose control flow crosses SIMT GPU execution and event-driven fabric execution.

\item We implement a prototype in Damer using dputime as the eBPF frontend and evaluate it on 1{,}000 generated movement microbenchmarks plus 14 LLM fabric workloads, showing where semantic policies improve modeled end-to-end data movement.
\end{itemize}

\section{Background and Motivation}
\label{sec:background}
\subsection{GPU Fabrics Are Already Cross-Layer}

Modern LLM systems are not only GPU compute pipelines.
They are data-movement systems whose critical paths cross GPU memory, host memory, CXL memory pools, RDMA queues, and collective communication libraries.
PagedAttention serving engines maintain logical-to-physical KV block tables~\cite{kwon2023vllm}.
NCCL and RDMA transports move activations, gradients, tokens, and checkpoints across GPUs~\cite{jeaugey2017nccl,hu2025demystifying}.
CXL 3.x adds fabric capabilities, switching, peer-to-peer access, memory sharing, and fabric manager mechanisms that make memory placement a fabric concern rather than a host-local concern~\cite{cxl2022specification,cxl2024specification}.
DPU and GPU-centric network APIs can remove the CPU from parts of the packet path, for example by enabling GPU-triggered network operations and direct GPU memory transfers~\cite{nvidiaDocaGpunetio,nvidiaDocaDpa}.

These mechanisms solve data-path problems.
They do not solve the policy problem.
The data path can move bytes efficiently once a transfer is chosen.
The difficult question is whether a transfer should be prioritized, delayed, transformed, redirected, replicated, persisted, or placed near the requester.
That decision needs execution semantics from the GPU side and fabric state from the network and CXL side.

\subsection{What Each Layer Can and Cannot See}

\paragraph{GPU view.}
GPU kernels and runtimes see the semantic structure of computation.
They know whether a stream is running prefill, decode, all-reduce, checkpoint, or MoE routing.
They can expose warp-level stalls, kernel phase, page faults, KV block reuse, sequence lifetime, expert ID, token batch, and whether a transfer is on the request critical path.
However, a GPU hook does not directly observe RDMA retransmissions, remote QP depth, CXL switch congestion, remote GPU readiness, or memory-pool pressure.

\paragraph{Fabric view.}
DPUs, NICs, CXL switches, and memory devices see fabric state.
They observe WQE and CQE timing, queue depth, retransmissions, timeouts, routing, memory-window ownership, persistence domains, and placement conflicts.
Some devices also provide programmable execution near the data path: BlueField DPA exposes event-triggered threads, asynchronous operations, RDMA primitives, memory services, and ordering plus notification control~\cite{nvidiaDocaDpa}; CXL switches and switch-attached near-memory engines provide a natural point for placement and copy policies when the traffic is CXL memory movement.
But the fabric sees a stream of operations, not their high-level meaning.
A QP does not reveal whether its bytes are decode KV blocks, MoE expert tokens, prefill activations, gradients, or checkpoint records.

\paragraph{Host view.}
The host CPU can coordinate runtimes, drivers, and fabric managers, but it is the wrong critical-path policy point for many decisions.
Host-mediated control adds latency, loses event granularity, and often observes state after queue buildup has already happened.
The host remains useful for admission control, policy loading, audit, and fallback.
\sys moves per-event decisions to the fabric while keeping the host as the policy owner.

\subsection{Why eBPF Is the Right Frontend}

eBPF provides a mature model for safe, dynamically loaded OS extensions.
It gives programmers maps, helpers, bounded execution, verifier checks, and attach points.
Prior work extended this model toward GPUs, showing that eBPF can expose GPU-side observability and policy hooks~\cite{yang2025egpu}.
\sys keeps eBPF as the frontend because fabric policies need the same properties: dynamic loading, per-tenant capability boundaries, bounded execution, and safe helper APIs.

The difference is scope.
A GPU-local eBPF policy can tune one accelerator.
\sys treats the fabric as the programmable object.
One logical policy is split into device-specific handlers:
\begin{itemize}[nosep]
\item GPU hooks publish semantic hints at kernel, stream, warp, page-fault, and phase boundaries.
\item Driver hooks bind streams, memory regions, QPs, CXL windows, and tenants to policy capabilities.
\item DPU/NIC hooks run at WQE submit, CQE complete, congestion update, timeout, and peer-failure events.
\item CXL switch or near-memory hooks run at placement, copy, persistence, reduction, and ownership-transfer events.
\end{itemize}

\subsection{From Shared Maps to Consistency Classes}

The tempting design is a shared cross-device BPF map.
That is insufficient.
Different policy state has different correctness requirements.
Approximate telemetry such as bytes moved or queue depth can be eventually consistent.
A decode hint can tolerate bounded staleness if it carries an epoch.
Ownership of a CXL memory window cannot be eventually consistent; a stale owner can violate isolation or ordering.
A retry command cannot be unbounded; it must have a TTL and exactly-once completion semantics.

\sys therefore treats map consistency as part of the policy type system.
The programmer chooses a consistency class, and the compiler/verifier checks which helpers may read or write that map from each device.
The runtime implements the chosen class with GPU-local maps, host mirrors, BlueField/DPU memory, CXL memory, or switch-local state depending on the target.

\subsection{Motivating Workloads}

\paragraph{LLM prefill attention and FFN.}
LLM prefill is the canonical \sys workload because the critical dataflow crosses all three execution points inside one request.
The attention path moves KV blocks, gathers remote cache lines, performs reductions, and depends on sequence-local readiness.
The FFN path moves activations and weights, often with shrinkable intermediates that can be quantized, compressed, filtered, or reduced before crossing the fabric.
GPU hooks expose the semantic phase and tensor objects; DPU/NIC hooks observe RDMA queues, congestion, and remote readiness; near-Type-2 CXL or switch-local hooks observe placement, DMA completion, and ownership beside the dataflow island.
No packet-only or page-only policy can see this full state.

\paragraph{MoE all-to-all.}
GPU hooks identify expert ID, token batch, source/destination rank, critical-path experts, and destination readiness.
Fabric hooks prioritize critical expert tokens, redirect tokens to idle replicas, pace congested experts, and avoid sending to a GPU whose receive buffer is not ready.

\paragraph{Prefill--decode colocation.}
When training, prefill, decode, and checkpoint traffic share the same GPU fabric, a QP-only policy is too coarse.
\sys classifies decode KV fetches as deadline-critical, prefill activations as medium priority, training all-reduce as paceable, and checkpoint traffic as deferrable under congestion.

\paragraph{Remote KV cache and CXL memory.}
GPU page-fault and KV-block hooks reveal reuse distance and sequence lifetime.
CXL switch or near-memory hooks can prefetch blocks, choose replicas, redirect failed memory servers, and place hot blocks near the GPU that will reuse them.

These examples are all data-movement policies.
They do not require inventing a new packet path or a new CXL protocol.
They require a safe way for execution semantics and fabric state to meet.

\section{Design}
\label{sec:method}
\subsection{Overview}

\sys compiles one logical fabric policy into a set of device-specific eBPF handlers and a shared movement plan.
The programmer writes policy over semantic events: GPU phase changes, page faults, stream boundaries, RDMA work-queue submission, completion events, congestion updates, CXL placement events, and ownership transfers.
The compiler lowers this policy into:
\begin{enumerate}[nosep]
\item a semantic movement graph;
\item per-device eBPF programs for GPU, driver, DPU/NIC, and CXL switch or near-memory hooks;
\item consistency-classed BPF maps and helper calls;
\item verifier obligations for bounded execution and cross-device effects; and
\item runtime manifests for dputime-based loading and observation.
\end{enumerate}

The design principle is simple: data-path mechanisms remain device-specific, but policy meaning is shared.
A DPU handler may post or defer an RDMA operation.
A CXL switch handler may choose placement or trigger a copy to a memory pool.
A GPU handler may publish decode or expert metadata.
All of them execute one policy contract with explicit consistency and capability rules.

\subsection{Semantic Movement Graph}

The compiler's internal representation is a movement graph.
Nodes describe where data resides or where transformations can execute:

\begin{center}
\small
\resizebox{\columnwidth}{!}{%
\begin{tabular}{ll}
\toprule
\textbf{Node} & \textbf{Meaning} \\
\midrule
host memory & CPU DRAM or pinned staging memory \\
CXL memory device & CXL Type-3 memory, pooled memory, or persistent memory \\
GPU memory & HBM or GPU-visible allocation \\
accelerator & GPU, DPU, or other compute-capable endpoint \\
switch compute engine & near-Type-2 small core plus local DMA/compute engines \\
\bottomrule
\end{tabular}
}
\end{center}

Edges describe a movement intent:
\begin{center}
\small
\resizebox{\columnwidth}{!}{%
\begin{tabular}{ll}
\toprule
\textbf{Field} & \textbf{Purpose} \\
\midrule
bytes, stride & data volume and access shape \\
reuse distance & whether placement or prefetch is useful \\
read/write ratio & copy, reduction, or writeback bias \\
source/destination & endpoints and routing candidates \\
ordering requirement & none, program order, acquire-release, or total \\
alias set & memory-equivalence and conflict class \\
ownership & borrowed, source-owned, destination-owned, or shared \\
transformation & operation fused with movement \\
\bottomrule
\end{tabular}
}
\end{center}

\sys currently recognizes nine movement actions:
\begin{center}
\small
\begin{tabular}{lll}
\toprule
Move & Move+Quantize & Move+Compress \\
Move+Checksum & Move+Filter & Move+Reduce \\
Move+Scatter/Gather & Move+Replicate & Move+Persist \\
\bottomrule
\end{tabular}
\end{center}

These actions are deliberately small.
They are expressive enough to cover common GPU-fabric transfers, but constrained enough for effect verification.
For example, a remote KV-cache policy may compile a page fault into prefetch-like behavior expressed as movement with replicate or persist semantics, depending on whether the target is a cache replica or durable CXL memory.
An MoE token routing policy compiles into scatter/gather movement plus optional filtering or replication.
Checkpoint and memory-tiering policies compile into movement with compression and persistence.

\subsection{Near-Type-2 Hardware JIT}

\sys separates fabric control from fabric dataflow.
The control point is a near-Type-2 small core: a programmable core near a CXL Type-2 endpoint, a switch-attached engine, or an equivalent accelerator-side control processor.
This core is not intended to copy every byte with scalar code.
Instead, it acts as a hardware-JIT for movement descriptors.
Given a verifier-approved edge, it specializes the descriptor into device-local commands: DMA copy, CXL placement, persistence, reduction, scatter/gather, pacing, or replica selection.
It also owns the local slice of policy state: map epochs, ownership handoff, command sequence numbers, TTLs, and capability checks.

The dataflow then runs in the surrounding Von Neumann island.
By this we mean the local memory, load/store engine, DMA engines, queue machinery, and small fixed-function or programmable transform units around the Type-2 endpoint.
The island still follows a conventional memory-command execution model, but \sys changes who programs it: the host no longer stages every policy decision, and the network no longer sees only packets or QPs.
The fabric small core turns semantic eBPF decisions into dataflow actions close to where CXL memory, GPU buffers, and switch-local resources meet.

Observation must be co-located with this execution point.
The dataflow is data-driven: work appears as queue entries, DMA descriptors, memory completions, ready bits, and ownership handoffs.
A host-side poller sees these signals late and after aggregation.
\sys therefore treats observation as a sidecar to the near-Type-2 hardware-JIT.
The small core observes local queue depth, completion timing, placement state, retry/redirect outcomes, and map-epoch transitions while it programs the local dataflow engines.
Those observations update telemetry and hint maps without moving the policy loop back to the host.

\subsection{LLM Prefill Across Three Architectures}

Prefill attention and FFN are the clearest example of why \sys is a fabric policy system rather than a DPU-only offload.
Within one prefill request, attention movement is dominated by KV-block lookup, gather, placement, reduction, and readiness; FFN movement is dominated by activation and weight movement plus shrinkable intermediate tensors.
\sys maps these data-driven flows across three architecture points:

\begin{center}
\small
\resizebox{\columnwidth}{!}{%
\begin{tabular}{p{0.24\columnwidth}p{0.34\columnwidth}p{0.34\columnwidth}}
\toprule
\textbf{Execution point} & \textbf{What it observes} & \textbf{What it controls} \\
\midrule
GPU SIMT/tensor execution & layer phase, tensor shard, KV block, page fault, stream deadline & HBM pre-transform, phase hints, block reuse, readiness publication \\
DPU/NIC event execution & WQE/CQE timing, QP depth, retransmit, congestion, remote GPU readiness & RDMA priority, pacing, redirect, retry, replica selection \\
near-Type-2 fabric island & CXL placement, DMA completion, queue state, ownership epoch, persistence state & hardware-JIT commands for copy, place, reduce, persist, replicate, and observe \\
\bottomrule
\end{tabular}
}
\end{center}

The same logical policy can therefore treat attention KV fetches as deadline-critical, pace FFN activation spill when queues build up, pre-transform shrinkable FFN intermediates on the GPU, and move cold KV or persistent state through a CXL-local dataflow island.
The observability point follows the dataflow: the GPU observes tensor semantics, the DPU observes network events, and the near-Type-2 core observes local queue and memory-state transitions.

\subsection{Policy Hooks}

\sys exposes hooks at the semantic points where information appears.
GPU hooks publish execution meaning; fabric hooks act on communication and memory-placement events.
The following example marks decode traffic at a GPU phase boundary:

\begin{lstlisting}
SEC("gpu/kernel_phase")
int mark_phase(struct gpu_kernel_ctx *ctx)
{
    struct fabric_hint hint = {
        .tenant   = ctx->tenant,
        .stream   = ctx->stream,
        .phase    = GPU_PHASE_DECODE,
        .deadline = ctx->deadline,
        .priority = LATENCY_CRITICAL,
    };

    bpf_fabric_publish(ctx->stream, &hint);
    return 0;
}
\end{lstlisting}

A fabric handler consumes that hint when a movement operation reaches the fabric:

\begin{lstlisting}
SEC("fabric/rdma_submit")
int schedule_wqe(struct fabric_wqe_ctx *ctx, struct fabric_hint *hint)
{
    if (hint && hint->phase == GPU_PHASE_DECODE)
        return BPF_FABRIC_HIGH_PRIORITY;
    return BPF_FABRIC_ALLOW;
}
\end{lstlisting}

The same policy may attach to multiple fabric hook kinds:
\begin{center}
\small
\resizebox{\columnwidth}{!}{%
\begin{tabular}{ll}
\toprule
\textbf{Hook} & \textbf{Typical decision} \\
\midrule
\texttt{rdma\_wqe\_submit} & priority, defer, redirect, pace \\
\texttt{rdma\_cqe\_complete} & update readiness, release ownership \\
\texttt{rx\_message}/\texttt{tx\_message} & packet or message classification \\
\texttt{congestion\_update} & pacing and replica choice \\
\texttt{remote\_gpu\_ready} & avoid CQ and buffer buildup \\
\texttt{flow\_timeout} & retry, failover, or TTL expiry \\
\texttt{peer\_failure} & route around failed GPU or memory server \\
\texttt{cxl\_place} & choose CXL memory device or switch path \\
\texttt{cxl\_copy\_complete} & publish placement epoch \\
\bottomrule
\end{tabular}
}
\end{center}

\subsection{Consistency-Classed BPF Maps}

The most important design choice is that \sys does not present one undifferentiated shared map.
Each map declares a consistency class:

\begin{figure}[t]
    \centering
    \includegraphics[width=\columnwidth]{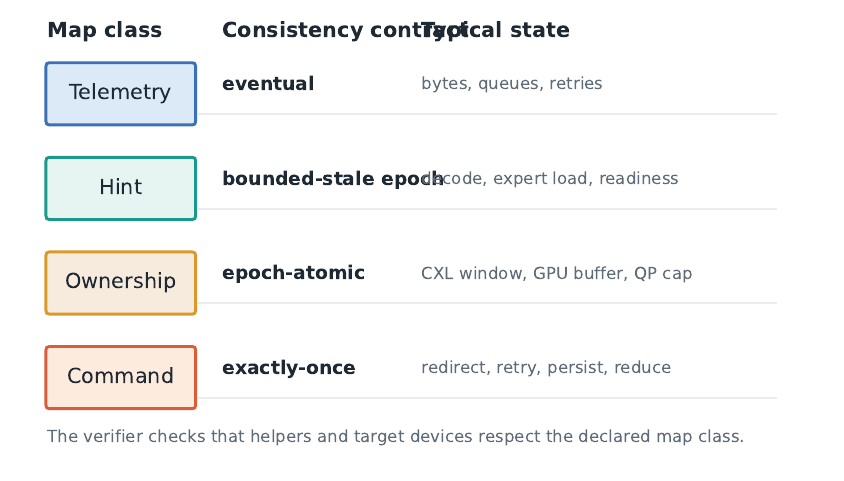}
    \caption{Consistency-classed BPF maps. Each map class exposes a different synchronization contract, allowing telemetry to remain cheap while ownership, placement, and commands receive stronger guarantees.}
    \label{fig:map-classes}
\end{figure}

\begin{center}
\small
\resizebox{\columnwidth}{!}{%
\begin{tabular}{p{0.25\columnwidth}p{0.66\columnwidth}}
\toprule
\textbf{Class} & \textbf{Contract} \\
\midrule
Telemetry & Eventually consistent counters and histograms. Used for bytes, queue depth, retry counts, or phase statistics. \\
Hint & Bounded-staleness values with an epoch and expiration. Used for priority, decode/prefill/MoE phase, expert load, and remote readiness. \\
Ownership & Epoch-atomic ownership or placement state. Used for CXL memory windows, GPU buffer ownership, QP capability binding, and alias sets. \\
Command & Exactly-once command completion with sequence numbers. Used for redirect, retry, persist, copy, reduce, and placement commits. \\
\bottomrule
\end{tabular}
}
\end{center}

The runtime implements these classes with hierarchical physical storage.
A GPU may keep a fast local map for phase hints.
The host mirror records policy ownership and audit state.
A DPU may keep hot QP and congestion entries in DPA memory.
A CXL switch or near-memory engine may keep placement and ownership epochs close to the memory pool.
The compiler generates synchronization code only where the class requires it.
Telemetry never pays ownership-map cost; ownership never degrades into best-effort statistics.

\subsection{Placement and Lowering}

\sys chooses where a movement-side transformation runs by considering source, destination, transform type, byte volume, reuse distance, ordering, and device capability.
The placement problem is not a global optimizer in the first prototype.
It is a constrained decision checked by the verifier:

\begin{figure}[t]
    \centering
    \includegraphics[width=\columnwidth]{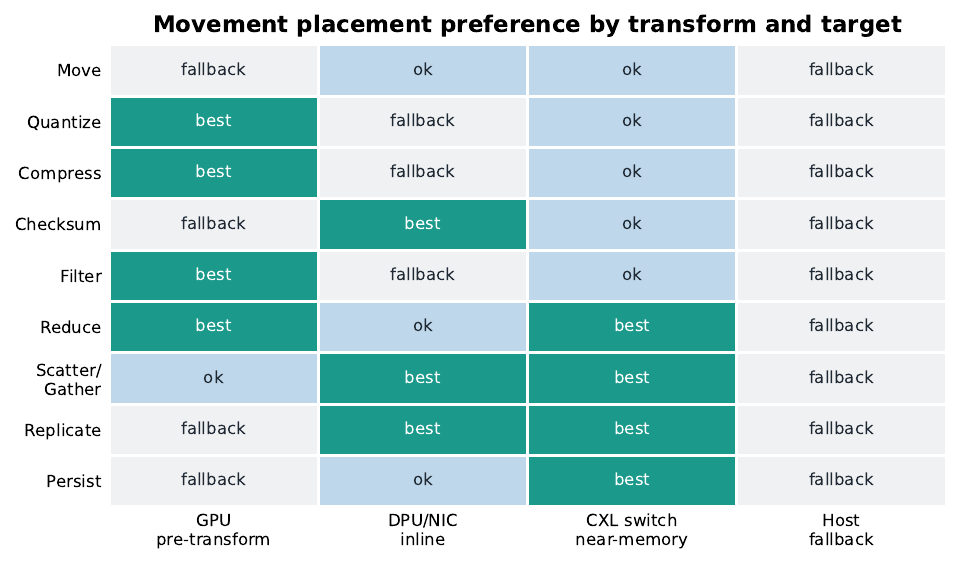}
    \caption{Placement preference for common movement transformations. GPU pre-transform is preferred when the transform shrinks HBM-resident data; DPU/NIC inline execution is preferred for network-path scheduling and copy control; CXL switch or near-memory execution is preferred for placement, persistence, and memory-local operations.}
    \label{fig:placement-matrix}
\end{figure}
\begin{itemize}[nosep]
\item GPU pre-transform is preferred when source data is in HBM and the transform shrinks data, such as quantize, compress, filter, or reduce.
\item DPU/NIC inline execution is preferred for RDMA queue scheduling, replica selection, pacing, checksum, and copy operations already on the network path.
\item CXL switch or near-memory execution is preferred for placement, persist, copy, reduce, and scatter/gather operations whose source or destination is CXL memory.
\item Host execution is a fallback for unsupported devices, policy loading, and audit.
\end{itemize}

For example, the following switchlet expresses a CXL-local quantizing move:

\begin{lstlisting}
damer.switchlet @kv_pack(
    %src : memref<?xf16, "cxl">,
    %dst : memref<?xi8, "cxl">) {
  %v = damer.read %src
  %q = damer.quantize %v
  damer.write %q, %dst
}
\end{lstlisting}

If the CXL memory device or switch-attached engine supports the transform, \sys keeps the operation near CXL memory.
If not, it may pre-transform on the GPU or fall back to host/DPU execution, but the verifier preserves ordering, alias, and ownership rules.

\subsection{Cross-Device Effect Verifier}

The verifier extends conventional eBPF bounded-execution checks with fabric effects.
It validates the combined policy, not only each handler in isolation.
The checked properties are:
\begin{enumerate}[nosep]
\item DPU, switch, and GPU handlers have bounded instruction count, helper count, and stack usage.
\item A handler cannot wait for a completion event that can only be produced by an operation the same handler has not yet issued.
\item GPU and fabric handlers cannot form a dependency cycle, such as GPU waiting for DPU completion, DPU waiting for a GPU-ready flag, and the GPU-ready flag being updated only after that completion.
\item Every QP, tenant, GPU memory region, and CXL window access is protected by a capability.
\item Each input event emits a bounded number of RDMA, DMA, or CXL memory operations.
\item Redirect and retry paths carry a TTL to prevent packet, WQE, or placement livelock.
\item GPU and fabric programs update through an epoch-atomic switch: either all devices execute the old policy or all execute the new policy for a given epoch.
\end{enumerate}

This is the core verification problem \sys targets: a single policy can have control flow that crosses SIMT GPU execution and event-driven fabric execution.
Without this verifier, semantic fabric policies become dangerous ad hoc offloads.
With it, \sys can allow dynamic policies while preserving tenant isolation and progress.

\section{Implementation}
\label{sec:implementation}
\sys is implemented in Damer as an eBPF middleware compiler.
The prototype is intentionally independent of CIRCT: eBPF is the frontend and runtime interface, while Damer owns the movement graph, placement decisions, helper ABI, verifier summary, and generated artifacts.

\subsection{Compiler Pipeline}

The compiler accepts two input forms.
First, a JSON event format describes movement requests emitted by GPU runtimes, drivers, or fabric monitors.
Second, a compact switchlet syntax describes high-level movement intent, as in the CXL-local \texttt{kv\_pack} example in Section~\ref{sec:method}.
Both are normalized into the same internal movement graph.

For each movement edge, the compiler emits:
\begin{itemize}[nosep]
\item a semantic plan JSON containing nodes, edges, placement, properties, and verifier results;
\item dputime-facing eBPF C with a policy section name and helper calls;
\item a dputime runtime manifest describing maps, helpers, and attach sections;
\item a BPF object compiled with \texttt{clang -target bpf}; and
\item an optional offload bundle for host/GPU/DPU observation.
\end{itemize}

The generated BPF C does not move user data itself.
It emits decisions and movement descriptors through two compact Damer helpers.
This keeps eBPF programs small and verifiable while leaving bulk transfer to RDMA, GPUDirect, CXL, DMA engines, or switch-local copy mechanisms.

\subsection{dputime Frontend}

\sys uses dputime as the userspace eBPF execution substrate and the dputime fork as the experimental GPU/DPU observation path.
On the GPU host, dputime can be built with CUDA attach support so CUDA module loading and runtime events become observable policy hooks.
On BlueField/Arm, dputime provides the native eBPF runtime used to observe DPU-side hooks and validate that generated BPF objects compile on the DPU operating system.

The runtime report records two observer roles:
\begin{center}
\small
\resizebox{\columnwidth}{!}{%
\begin{tabular}{ll}
\toprule
\textbf{Observer} & \textbf{Role} \\
\midrule
GPU host & observe CUDA/GPU-side policy hooks and memory movement intent \\
BlueField DPU & observe RDMA/NIC placement, queue events, and offload bundle state \\
CXL switch target & consume movement plans for switch or near-memory execution \\
\bottomrule
\end{tabular}
}
\end{center}

The third target is currently a backend interface rather than a vendor-specific switch firmware implementation.
The compiler represents it as \texttt{switch\_compute\_engine} and emits placement, consistency, and verifier metadata that a CXL switch-attached executor or near-memory controller can consume.
This is deliberate: \sys should not bake BlueField DPA into the abstraction boundary.
DPUs are one execution point; CXL fabric points are another.
For the CXL target, the emitted artifact is a hardware-JIT input rather than a monolithic kernel: it contains a verified movement descriptor, map-class requirements, command-map completion rules, and the local state that a near-Type-2 small core needs to program the adjacent dataflow engines.
It also contains observer metadata for data-driven events next to those engines, including queue updates, DMA completion, memory-placement completion, ownership handoff, and retry or redirect expiry.

\subsection{Runtime Artifacts}

An offload bundle contains all state needed to load and audit a policy:
\begin{itemize}[nosep]
\item the source workload event or switchlet;
\item generated movement plans;
\item BPF C and BPF objects;
\item dputime manifests;
\item verifier summaries;
\item helper IDs and map declarations;
\item target roles and observer build information; and
\item policy epoch metadata.
\end{itemize}

This bundle is the unit of cross-device update.
When a policy is installed, each target first loads the new program into an inactive epoch.
Only after all required targets acknowledge verification and load success does the host publish the new epoch.
Fabric events then carry an epoch tag, so a GPU hint, a DPU queue decision, and a CXL placement decision agree on which policy version they are executing.

\subsection{Helper ABI}

\sys exposes helper APIs at the movement level rather than at raw packet or page granularity.
Table entries below use short names; generated code uses the corresponding \texttt{bpf\_fabric\_*} or \texttt{bpf\_cxl\_*} symbol.
Important helpers include:

\begin{center}
\small
\resizebox{\columnwidth}{!}{%
\begin{tabular}{p{0.36\columnwidth}p{0.55\columnwidth}}
\toprule
\textbf{Helper} & \textbf{Purpose} \\
\midrule
\texttt{publish} & publish a semantic hint with epoch and TTL \\
\texttt{lookup} & read a hint or ownership record under its map class \\
\texttt{ready} & query destination readiness before enqueueing movement \\
\texttt{emit} & emit a movement decision descriptor \\
\texttt{redirect} & choose an alternate QP, GPU, replica, or CXL target \\
\texttt{cxl\_place} & request CXL memory placement or migration \\
\texttt{commit} & commit completion for exactly-once command maps \\
\bottomrule
\end{tabular}
}
\end{center}

The verifier checks helper use against map class and target capability.
For example, a GPU hook may publish a hint but may not directly commit ownership of a CXL memory window unless it holds the corresponding capability.
A CXL placement hook may update an ownership map only through an epoch-checked helper.
A DPU handler may emit a bounded number of RDMA operations and must attach TTL to redirect or retry decisions.

\subsection{CXL Switch and Near-Memory Backend}

\sys treats CXL as both a memory target and a policy target.
When an edge's source or destination is a CXL-memory node, the compiler considers CXL-local placement.
When a transform is bandwidth-reducing or persistence-related, it may place the transform at a switch compute engine if the target capability exists.
The intended backend is a near-Type-2 control core that performs hardware-JIT specialization and state management, then dispatches the operation to the local Von Neumann island: CXL memory, queues, DMA, and simple transform engines.
Observation is placed next to the same island.
The backend records state changes where the dataflow is produced, not only after the host observes a completed transfer.

The first prototype models this backend with the same movement-plan interface used for DPU and GPU targets.
The plan records:
\begin{itemize}[nosep]
\item the selected CXL memory device or pool;
\item whether the operation requires program order, acquire-release, or total ordering;
\item alias and ownership epochs;
\item expected byte volume after transformation;
\item data-driven observation points beside queue, DMA, and placement engines;
\item fallback targets if switch-local execution is unavailable; and
\item command-map completion conditions.
\end{itemize}

This lets the evaluation ask a concrete question before vendor-specific switch support is available: where would a near-Type-2 hardware-JIT and its local dataflow island reduce end-to-end movement, and which events are still dominated by GPU or DPU paths?

\subsection{Failure and Fallback}

\sys assumes policy handlers are fail-stop and bounded by the verifier.
If a target cannot load a policy, the epoch switch is aborted and the previous policy remains active.
If a target disappears at runtime, capability checks force dependent events onto fallback paths.
For example, if the CXL switch-local executor is unavailable, Move+Persist can fall back to DPU inline copy or host-mediated placement, but ownership epochs still prevent two devices from claiming the same memory window.
If the DPU runtime is unavailable, GPU and host hooks can continue publishing hints and using conservative default fabric decisions.

\section{Evaluation}
\label{sec:evaluation}
We evaluate \sys with three goals.
First, can the compiler generate verified eBPF artifacts across the movement primitive space?
Second, can the same workload be compiled into host/GPU/DPU-facing artifacts rather than remaining a paper-only abstraction?
Third, where does semantic fabric policy improve modeled end-to-end data movement, especially when CXL memory and switch-local placement are considered?

\subsection{Experimental Setup}

\paragraph{Prototype.}
The prototype is implemented in Damer.
It compiles movement events and switchlets into semantic plans, BPF C, BPF objects, and runtime manifests.
The dputime fork of dputime is used as the GPU/DPU observation runtime.
The GPU-host path is built with CUDA attach support when CUDA headers and a visible NVIDIA GPU are available.
The BlueField path is built natively on Arm Ubuntu and validates BPF compilation on the DPU OS.
The CXL switch backend is evaluated through the \texttt{switch\_compute\_engine} placement model and verifier metadata because commodity programmable CXL-switch firmware is not part of the prototype.

\paragraph{Workloads.}
We use 14 fabric workloads:
\begin{itemize}[nosep]
\item \textbf{Qwen-style serving}: logits movement, KV packing, prefill activation spill, decode KV fetch, tensor shard exchange, KV replication, and checkpoint persistence.
\item \textbf{Qwen DPU decode}: decode-critical KV fetch, token scatter, retry/redirect, replica refresh, and append-log persist decisions visible to the DPU.
\item \textbf{Qwen long-context CXL}: CXL-backed KV fetch, prefix-cache replication, cold-KV eviction, sequence migration, and attention reduction.
\item \textbf{Qwen parallelism}: pipeline activation movement, CXL activation spill, tensor all-reduce, weight streaming, microbatch scatter, and checkpoint persist.
\item \textbf{Qwen LoRA adapter}: adapter delta reduction, quantized adapter push, optimizer-state compression, decode KV fetch, adapter replica refresh, and checkpoint persist.
\item \textbf{Qwen scale-out}: 2/4/8/16/32 logical GPU variants with increasing collective, scatter/gather, replica, redirect, and checkpoint fanout.
\item \textbf{MoE all-to-all}: critical expert tokens, background expert traffic, expert output reduction, replica refresh, and checkpoint persist.
\item \textbf{Prefill/decode colocation}: decode KV fetch, prefill activation spill, background training all-reduce, checkpoint throttle, and shared tensor movement.
\item \textbf{Remote KV cache}: KV prefetch, compact/persist, failed memory-server redirect, sequence migration, and placement updates.
\item \textbf{Training/checkpoint}: gradient reduction, parameter broadcast, optimizer shard exchange, activation checkpoint compression, and step checkpoint persist.
\end{itemize}

\paragraph{Metrics.}
For artifact-level evaluation we report generated cases, verifier success, BPF object count, and compiler latency.
For E2E evaluation we use the explicit data-movement model recorded in the benchmark report.
The baseline is staged host/NIC-visible movement with separate transform and control steps.
The \sys path uses fused placement and dputime offload control.
A second optimizer evaluates copy-path choices on top of \sys: DPU inline, GPU pre-transform, and GPU peer/CXL direct paths.
These are modeled data-movement results, not application token-latency measurements.

\subsection{Microbenchmark Coverage}

The generated microbenchmark suite spans all nine movement kinds and five node classes.
It validates that the compiler can parse events, build movement plans, run verifier checks, emit BPF C, and produce bounded artifacts.

\begin{table}[t]
\centering
\caption{Generated movement microbenchmark coverage.}
\label{tab:microbench}
\begin{tabular}{lr}
\toprule
\textbf{Metric} & \textbf{Value} \\
\midrule
Cases & 1{,}000 \\
Movement kinds & 9 \\
Node classes & 5 \\
Placement targets & 5 \\
Max RDMA/CXL ops per event & 8 \\
Median compile time & 8.5\,$\mu$s \\
95th percentile compile time & 9.7\,$\mu$s \\
Median BPF C emit time & 4.1\,$\mu$s \\
\bottomrule
\end{tabular}
\end{table}

Table~\ref{tab:microbench} shows that the compiler path is lightweight enough for policy iteration.
The point is not that policy compilation occurs on every packet or WQE.
Rather, fabric operators can generate, verify, and load specialized movement policies at runtime without treating the compiler as an offline-only tool.

\subsection{Verifier Coverage}

We evaluate the verifier as a compiler component: a test passes when an input with a legal effect contract is accepted, or when an input with an illegal effect contract is rejected with a diagnostic.
Table~\ref{tab:verifier-coverage} summarizes positive, fuzzed, and targeted negative coverage.
The fuzzer mixes JSON events, compact switchlet syntax, and multi-edge pipelines.
The targeted negative suite exercises the implemented hard effect checks: transform-count bounds, RDMA/CXL fanout bounds, and retry/redirect TTL requirements.

\begin{table}[t]
\centering
\caption{Verifier regression coverage. Expected rejections count as passing tests when the verifier rejects the input.}
\label{tab:verifier-coverage}
\small
\resizebox{\columnwidth}{!}{%
\begin{tabular}{llrl}
\toprule
\textbf{Suite} & \textbf{Expected result} & \textbf{Pass rate} & \textbf{Coverage} \\
\midrule
Generated positives & Accept & 1{,}000/1{,}000 & 9 actions, 5 nodes, 5 placements; max fanout 8 \\
LLM workload events & Accept & 86/86 & 14 workloads, 10 movement kinds, 3 placements; max fanout 64 \\
Fuzzer mix & Accept/reject & 512/512 & 404 accepted, 108 rejected, 0 unexpected; 204 kinds, 40 multi-edge \\
Targeted negatives & Reject & 3/3 & transform bound, RDMA/CXL fanout bound, retry TTL \\
\bottomrule
\end{tabular}
}
\end{table}

The LLM workloads produce no verifier diagnostics.
Each accepted plan carries a bounded-effect summary; in this suite the generated helper bound is two helper calls per event and the largest event fanout is 64, below the configured bound of 128.
The targeted negative cases are intentionally small, because their purpose is not random discovery but regression protection for specific safety obligations.

\subsection{End-to-End Workload Compilation}

Across the 14 workloads, \sys compiles 86 semantic movement events into 86 BPF objects.
The same artifact bundle is shipped to the BlueField host for remote Arm compilation; all 86 generated BPF C programs recompile successfully on the DPU OS.
This validates the cross-device packaging path: the generated policy is not only a local plan but a set of BPF artifacts that can be observed by host and DPU runtimes.

\begin{table}[t]
\centering
\caption{Modeled data-movement E2E speedup over staged host/NIC control.}
\label{tab:e2e}
\small
\resizebox{\columnwidth}{!}{%
\begin{tabular}{lrrr}
\toprule
\textbf{Workload} & \textbf{Events} & \textbf{\sys us} & \textbf{Speedup} \\
\midrule
Qwen-style serving & 9 & 10{,}146.8 & 2.77$\times$ \\
Qwen DPU decode & 6 & 3{,}780.3 & 2.79$\times$ \\
Qwen long-context CXL & 6 & 21{,}884.2 & 4.10$\times$ \\
Qwen pipeline/tensor parallel & 6 & 52{,}957.2 & 3.06$\times$ \\
Qwen LoRA adapter & 6 & 9{,}586.4 & 3.65$\times$ \\
Qwen scale-out 2 GPU & 6 & 12{,}965.2 & 2.76$\times$ \\
Qwen scale-out 4 GPU & 6 & 27{,}236.4 & 2.66$\times$ \\
Qwen scale-out 8 GPU & 6 & 59{,}973.2 & 2.50$\times$ \\
Qwen scale-out 16 GPU & 6 & 142{,}224.1 & 2.26$\times$ \\
Qwen scale-out 32 GPU & 6 & 373{,}834.6 & 1.96$\times$ \\
MoE all-to-all & 5 & 2{,}394.0 & 3.90$\times$ \\
Prefill/decode colocation & 6 & 18{,}616.8 & 2.87$\times$ \\
Remote KV cache/CXL memory & 6 & 3{,}621.4 & 3.84$\times$ \\
Training/checkpoint & 6 & 71{,}159.5 & 2.56$\times$ \\
\midrule
Aggregate & 86 & 810{,}380.1 & 2.34$\times$ \\
\bottomrule
\end{tabular}
}
\end{table}

Table~\ref{tab:e2e} reports the first-level \sys speedup.
The largest relative gains appear in long-context CXL, MoE, remote KV cache, and LoRA adapter workloads because semantic routing, placement, and transformation reduce unnecessary movement and queueing.
The largest absolute latency remains in training/checkpoint, pipeline/tensor parallel, and scale-out traffic because those events carry large byte volumes.
This distinction is important: semantic policies help critical small transfers and large background transfers for different reasons.

\subsection{Qwen Scale-Out}

\begin{figure}[t]
    \centering
    \includegraphics[width=\columnwidth]{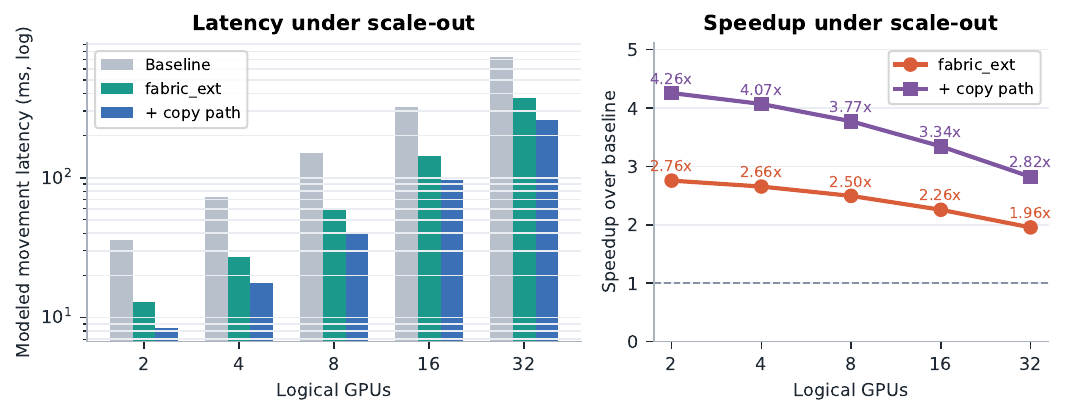}
    \caption{Qwen 27B scale-out experiment. Labels show first-level \sys speedup and speedup after copy-path optimization.}
    \label{fig:qwen-scaleout}
\end{figure}

The scale-out workloads increase tensor collective size, scatter/gather fragments, KV replica fanout, redirect fanout, and checkpoint bytes from 2 to 32 logical GPUs.
Figure~\ref{fig:qwen-scaleout} shows that \sys preserves a 1.96--2.76$\times$ modeled speedup as fanout grows.
The decline at 16 and 32 GPUs is expected: replica refresh and checkpoint persistence become byte-volume dominated, so semantic policy cannot hide all fabric work.
Copy-path optimization still adds 1.44--1.54$\times$, primarily by selecting DPU inline replica and checkpoint persistence plus direct GPU-peer scatter/gather paths.

\subsection{Where Copy-Path Optimization Matters}

We next ask where further DPU/GPU/CXL copy-path optimization helps on top of the current \sys placement.
The optimizer considers three additional candidates:
\begin{itemize}[nosep]
\item \textbf{DPU inline}: keep copy or persistence control on the BlueField/GPUDirect side and reduce staged control overhead.
\item \textbf{GPU pre-transform}: perform shrinking transformations such as compress, quantize, filter, or reduce on the GPU before fabric movement.
\item \textbf{GPU peer/CXL direct}: use direct GPU--GPU or GPU--CXL paths when the edge does not need host staging.
\end{itemize}

\begin{table}[t]
\centering
\caption{Copy-path optimization on top of \sys.}
\label{tab:copyopt}
\small
\resizebox{\columnwidth}{!}{%
\begin{tabular}{lrrr}
\toprule
\textbf{Workload} & \textbf{Current us} & \textbf{Optimized us} & \textbf{Extra speedup} \\
\midrule
Qwen-style serving & 10{,}146.8 & 6{,}865.3 & 1.48$\times$ \\
Qwen DPU decode & 3{,}780.3 & 2{,}504.4 & 1.51$\times$ \\
Qwen long-context CXL & 21{,}884.2 & 10{,}523.1 & 2.08$\times$ \\
Qwen pipeline/tensor parallel & 52{,}957.2 & 35{,}618.1 & 1.49$\times$ \\
Qwen LoRA adapter & 9{,}586.4 & 5{,}269.2 & 1.82$\times$ \\
Qwen scale-out 2 GPU & 12{,}965.2 & 8{,}398.3 & 1.54$\times$ \\
Qwen scale-out 4 GPU & 27{,}236.4 & 17{,}772.1 & 1.53$\times$ \\
Qwen scale-out 8 GPU & 59{,}973.2 & 39{,}665.3 & 1.51$\times$ \\
Qwen scale-out 16 GPU & 142{,}224.1 & 96{,}034.6 & 1.48$\times$ \\
Qwen scale-out 32 GPU & 373{,}834.6 & 259{,}105.0 & 1.44$\times$ \\
MoE all-to-all & 2{,}394.0 & 895.0 & 2.67$\times$ \\
Prefill/decode colocation & 18{,}616.8 & 12{,}779.4 & 1.46$\times$ \\
Remote KV cache/CXL memory & 3{,}621.4 & 1{,}943.9 & 1.86$\times$ \\
Training/checkpoint & 71{,}159.5 & 49{,}300.0 & 1.44$\times$ \\
\midrule
Aggregate & 810{,}380.1 & 546{,}673.8 & 1.48$\times$ \\
\bottomrule
\end{tabular}
}
\end{table}

Table~\ref{tab:copyopt} shows that copy-path optimization raises aggregate modeled speedup from 2.34$\times$ to 3.47$\times$ over the staged baseline.
The path mix is 49 DPU inline choices, 24 GPU pre-transform choices, and 13 direct GPU-peer/CXL choices.
Critical-priority events improve by 1.51$\times$ over the current \sys path.

\begin{table}[t]
\centering
\caption{Largest copy-path savings over current \sys placement.}
\label{tab:hotspots}
\small
\resizebox{\columnwidth}{!}{%
\begin{tabular}{llr}
\toprule
\textbf{Event} & \textbf{Best path} & \textbf{Saved} \\
\midrule
Qwen 32-GPU KV replica refresh & DPU inline & 45.33\,ms \\
Qwen 32-GPU checkpoint persist & DPU inline & 31.32\,ms \\
Qwen 32-GPU tensor scatter & GPU peer/CXL direct & 25.08\,ms \\
Qwen 16-GPU checkpoint persist & DPU inline & 15.66\,ms \\
Qwen 16-GPU tensor scatter & GPU peer/CXL direct & 12.54\,ms \\
Qwen 16-GPU KV replica refresh & DPU inline & 11.48\,ms \\
Qwen 32-GPU tensor all-reduce & GPU pre-transform & 8.95\,ms \\
Qwen long-context cold KV evict & GPU pre-transform & 8.59\,ms \\
\bottomrule
\end{tabular}
}
\end{table}

Table~\ref{tab:hotspots} answers the practical optimization question.
Large persistent or replicated transfers benefit most from DPU inline copy control.
GPU-source compress and reduce events benefit from pre-transform because fewer bytes cross PCIe, RDMA, or CXL.
MoE all-to-all benefits from direct peer routing, but its absolute savings are smaller unless expert traffic is on the critical path.

\begin{figure}[t]
    \centering
    \includegraphics[width=\columnwidth]{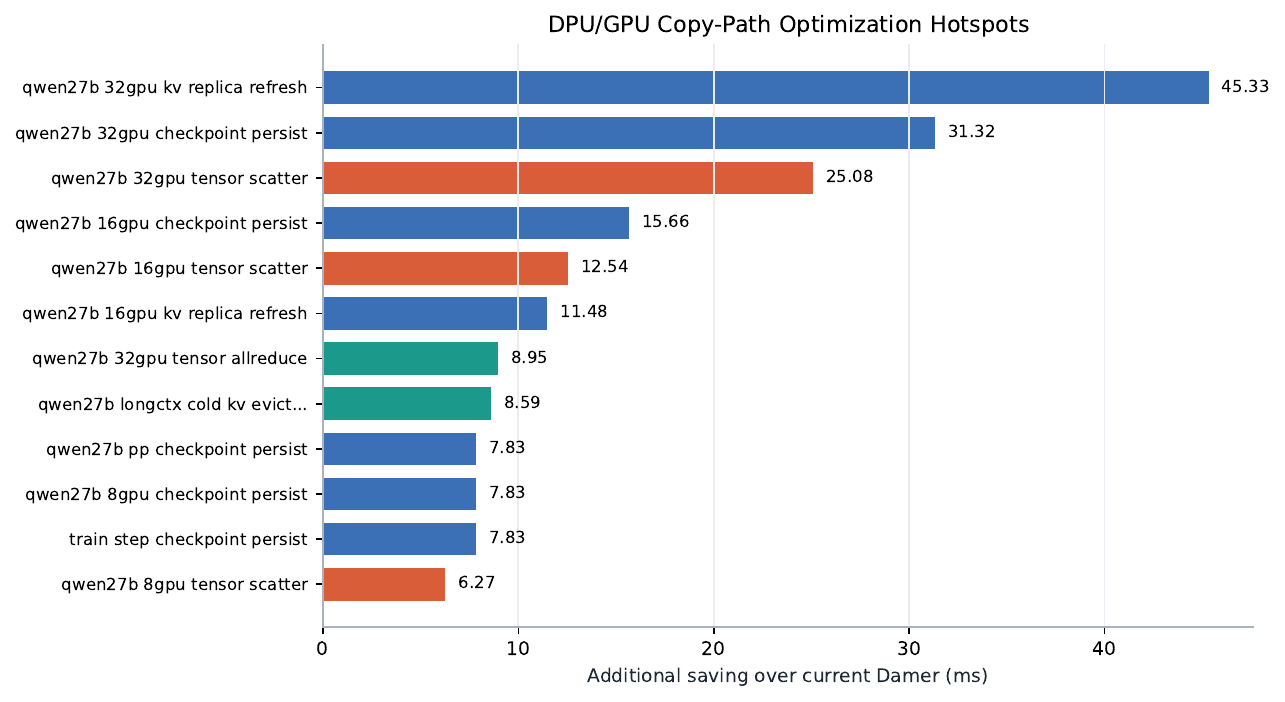}
    \caption{Event-level copy-path optimization hotspots. The largest absolute E2E savings come from DPU inline persist/replicate paths and GPU-side pre-transform for compress/reduce traffic.}
    \label{fig:copy-hotspots}
\end{figure}

\subsection{Case Studies}

\paragraph{LLM prefill attention and FFN.}
Prefill is the most complete use case because attention and FFN exercise different parts of the same fabric policy.
Attention KV fetches and reductions benefit from GPU semantic hints plus DPU/NIC readiness and pacing.
Long-context KV placement and cold-block movement benefit from the CXL or switch-local target because ownership and placement are memory-side decisions.
FFN activation spill and tensor-parallel exchange benefit from GPU pre-transform when intermediates shrink, and from DPU inline control when the edge is already on the RDMA path.
This is exactly the three-architecture split \sys is designed for: GPU tensor execution names the data, DPU/NIC event execution schedules network movement, and the near-Type-2 island JITs and observes local dataflow commands.

\paragraph{Semantic MoE scheduling.}
The MoE workload has the highest extra speedup after copy-path optimization.
Critical expert tokens use high-priority fabric decisions, while background expert traffic can use GPU-peer/CXL-direct movement.
The policy is semantic: it does not merely prioritize a QP, but ties fabric scheduling to expert ID, token criticality, and destination readiness.

\paragraph{Prefill/decode colocation.}
Decode KV fetches are small but deadline-sensitive, so their improvement is mostly queueing and readiness.
Checkpoint throttle and prefill activation spill are larger, so their improvement comes from DPU inline copy and GPU-side compression.
This validates the need for multiple policy classes within one workload.

\paragraph{Remote KV over CXL memory.}
Remote KV cache events combine page-fault semantics from the GPU with placement and persistence decisions in the CXL fabric.
The largest gains come from compact/persist and failed-server redirect paths.
This is the main example where CXL switch or near-memory placement is the correct policy target: the decision is about memory location and ownership, not only about RDMA priority.

\subsection{Takeaways}

The evaluation supports three claims.
First, the eBPF middleware compiler can cover the movement primitive space and produce real BPF artifacts.
Second, semantic policies expose E2E effects that packet/QP-only policies miss, especially MoE routing, decode/prefill colocation, and remote KV placement.
Third, the right execution point varies by event.
DPU inline execution is best for large copy and persist paths; GPU pre-transform is best when HBM-side shrinking reduces fabric bytes; CXL switch or near-memory execution is best when the policy is placement, ownership, persistence, or memory-local reduction.

\section{Related Work}
\label{sec:related}
\paragraph{eBPF beyond the kernel.}
eBPF has become a practical substrate for safe OS extension because it combines dynamic loading, helper APIs, maps, and verifier-enforced bounded execution.
Recent work has explored eBPF deployment outside the kernel and into heterogeneous environments, including GPU observability and programmability~\cite{yang2025egpu}.
\sys builds on this direction but changes the unit of programmability.
Rather than programming one GPU or one host kernel hook, it programs a fabric policy that spans GPU execution, driver binding, DPU/NIC events, and CXL placement.
The main new mechanism is the cross-device contract: consistency-classed maps and a verifier for effects that cross device boundaries.

\paragraph{GPU runtime and cluster scheduling.}
Systems such as Salus, Tiresias, Gavel, and SIA schedule GPU clusters using host-visible metrics and workload-level objectives~\cite{yu2020salus,gu2019tiresias,narayanan2020gavel,subramanya2023sia}.
They are complementary to \sys.
A cluster scheduler decides which job or model should receive resources.
\sys decides, at fabric-event granularity, how semantically different transfers within a job should move through GPU, RDMA, and CXL paths.
For example, a scheduler may co-locate prefill and decode on a GPU pool; \sys can still prioritize decode KV fetches over checkpoint writes inside that shared fabric.

\paragraph{GPU communication and RDMA.}
NCCL and RDMA transports provide high-throughput GPU communication~\cite{jeaugey2017nccl,hu2025demystifying}.
DOCA GPUNetIO and related GPU-centric networking mechanisms reduce CPU involvement in packet and RDMA paths~\cite{nvidiaDocaGpunetio}.
BlueField DPA provides event-triggered execution, RDMA primitives, memory services, and ordering plus notification control near the network path~\cite{nvidiaDocaDpa}.
\sys does not replace these mechanisms.
It supplies a policy layer above them, allowing scheduling and placement decisions to use GPU execution semantics instead of only QP, packet, or byte-level information.

\paragraph{CXL and disaggregated memory.}
CXL enables memory expansion, pooling, switching, peer-to-peer access, and fabric management~\cite{cxl2022specification,cxl2024specification}.
Systems such as DirectCXL, Pond, TPP, and related tiered-memory work study how to expose and manage CXL memory from hosts and applications~\cite{gouk2022direct,wang2022pond,maruf2023tpp}.
\sys is not a CXL memory allocator by itself.
It is a policy compiler that can use CXL memory and switch-local execution as targets when a GPU workload exposes semantic movement intent such as KV block reuse, ownership, persistence, or reduction.

\paragraph{LLM serving systems.}
LLM serving engines such as vLLM expose structured memory behavior through KV block tables and PagedAttention~\cite{kwon2023vllm}.
This structure is exactly the kind of semantic signal \sys wants to carry into the fabric.
The difference is boundary.
Serving engines usually optimize within the GPU runtime or application scheduler.
\sys lets those signals influence RDMA, DPU, and CXL decisions without hard-coding every policy into the serving engine.

\paragraph{Near-data and switch execution.}
Near-data processing has long argued that computation should move toward data when movement dominates cost.
\sys applies the same idea to fabric policies rather than general-purpose computation.
It does not require arbitrary application kernels to run inside a switch.
Instead, it targets small bounded movement-side operations---copy, persist, reduce, checksum, filter, replicate, and placement commit---that are suitable for verification and execution near RDMA or CXL data paths.

\section{Discussion}
\label{sec:discussion}
\subsection{Why Not a DPU-Only Design?}

A DPU is a powerful policy point, but it is not the fabric.
DPU/NIC hooks are ideal for RDMA queue scheduling, retransmission response, pacing, and network-path copy control.
They are the wrong place to observe warp stalls, kernel phase, KV block lifetime, and GPU page faults.
They are also not always the closest place to CXL memory placement or persistence.
\sys therefore treats the DPU as one executor among several.
The abstraction boundary is the semantic movement graph plus verified policy contract, not a particular BlueField generation or a particular NIC API.

\subsection{Why Not Just Use a Fabric Manager?}

CXL fabric managers configure routes, devices, memory windows, and topology.
They are necessary, but they usually operate at management timescales and with host-visible information.
\sys targets per-event policy decisions inside running GPU workloads.
For example, a fabric manager can make a CXL memory pool available; \sys can decide that a specific remote KV block should be prefetched, replicated, or persisted because the GPU just exposed its reuse distance and sequence lifetime.
The two systems are complementary: the fabric manager owns topology and administrative policy, while \sys handles verified movement decisions within those boundaries.

\subsection{Policy Stability and Debugging}

Cross-device policies are harder to debug than local eBPF programs.
\sys addresses this by making policy state explicit.
Every generated artifact records helper IDs, map classes, capability requirements, verifier summaries, placement decisions, and policy epochs.
Telemetry maps are intentionally cheap and approximate so operators can observe policy behavior without perturbing ownership or command state.
For stronger classes, \sys records epoch transitions and command completions so policy changes can be audited after a failure or performance anomaly.

\subsection{Security and Isolation}

The main isolation risk is that a semantic policy might affect resources outside its tenant or memory region.
\sys uses capabilities at every cross-device boundary.
A handler cannot access an RDMA QP, GPU region, CXL window, tenant hint, or ownership map unless the loader grants the capability.
Map class also matters for security.
Telemetry writes cannot mutate ownership.
Hints expire and carry epochs.
Ownership and command maps require checked helpers and cannot be updated through raw map stores.

\subsection{Limitations}

The current implementation is a prototype.
It generates real BPF artifacts and validates host/GPU/DPU packaging, but the near-Type-2 hardware-JIT backend is modeled through the movement-plan interface rather than deployed on a production programmable CXL switch or Type-2 device.
The E2E speedups in Section~\ref{sec:evaluation} are modeled data-movement speedups, not token-latency measurements from a full serving stack.
This distinction is important: the results identify where semantic policy should improve end-to-end movement, but real deployment will add device firmware constraints, runtime integration cost, and workload-specific scheduling effects.

\sys also assumes fail-stop behavior for policy handlers.
The verifier bounds execution and effects, but it does not prove arbitrary semantic correctness of user policy.
A bad but bounded policy can still choose poor priorities.
Finally, some fabric operations require vendor-specific attach points.
\sys's contribution is the common policy contract and compiler structure; each hardware target still needs a backend that maps verified decisions to device-specific mechanisms.

\subsection{Future Work}

Three directions are immediate.
First, a production CXL/Type-2 backend should map \texttt{switch\_compute\_engine} placement decisions to a hardware-JIT interface on the fabric small core and to real switch, Type-2, or near-memory firmware APIs.
Second, the copy-path optimizer should be connected to hardware counters from GPUs, DPUs, and CXL devices so modeled bandwidth and queue costs become measured online values.
Third, higher-level LLM runtimes should expose richer semantic hints---expert replica health, KV block temperature, sequence deadlines, and decode batch structure---so \sys can make better fabric decisions without reverse-engineering application state.

\section{Conclusion}
\label{sec:conclusion}
\sys argues that the programmable object in modern GPU systems is the fabric, not an individual device.
GPU kernels expose execution semantics that the network and CXL fabric cannot infer from packets or memory transactions.
DPUs, NICs, and CXL switches expose congestion, placement, ordering, and ownership state that GPU-local policies cannot see.
Treating these as separate control systems leaves performance and correctness decisions at the wrong layer.

\sys provides a cross-layer eBPF middleware compiler for this setting.
It represents movement as a semantic graph, lowers one logical policy into GPU, driver, DPU/NIC, and CXL switch or near-memory hooks, and protects cross-device state with consistency-classed BPF maps.
Its verifier checks bounded execution, bounded fabric effects, capability access, retry TTLs, dependency cycles, and epoch-atomic program updates across devices.
At the fabric edge, the key execution point is a near-Type-2 small core that performs hardware-JIT specialization, state management, and adjacent observation, while the surrounding Von Neumann island executes data-driven movement dataflow.
LLM prefill attention and FFN form the canonical use case: one request exposes tensor semantics on the GPU, network pressure on the DPU/NIC, and placement plus ownership transitions at the CXL or switch-local island.

The prototype shows that this design can generate real dputime-facing artifacts and expose meaningful E2E optimization points.
Across 14 LLM fabric workloads and 86 generated BPF objects, semantic placement and offload reduce modeled data-movement time by 2.34$\times$ over staged host/NIC control; copy-path selection across DPU inline, GPU pre-transform, and direct GPU-peer/CXL paths raises the modeled speedup to 3.47$\times$.
The central lesson is that the fabric can make better decisions when it sees GPU semantics, and GPU policies become more effective when they can safely act on fabric state.

\bibliographystyle{plain}
\bibliography{ref}

\end{document}